# Impact of Electrical Contacts on Transition Metal Dichalcogenides-Based Acoustoelectric and Acousto-Photoelectric Devices

*Benjamin Mayer[1], Felix M. Ehring[1], Clemens Strobl[1], Matthias Weiß[1], Ursula Wurstbauer[1,2], Hubert J. Krenner[1] and Emeline D. S. Nysten[1,\*]*

[1]Physikalisches Institut, Universität Münster, Wilhelm-Klemm-Straße 10, 48149 Münster, Germany

[2]Center for Soft Nanoscience (SoN), Universität Münster, Busso-Peus-Str.10, 48149 Münster

ABSTRACT

We study the impact of electrical contact barriers in hybrid $WSe_2$-$LiNbO_3$-based acoustoelectric and acousto-photoelectric devices using a combination of scanning photocurrent and acousto-electric current spectroscopy. Static scanning photocurrent measurements provide a qualitative measure of the band-bending and spatial distribution of the Schottky barrier between semiconducting $WSe_2$ and gold electrodes whereas the surface acoustic wave- induced acousto-electric current reveals the height of the tunnelling barrier created by the van der Waals gap between the multilayered $WSe_2$ flake and the gold electrodes in addition to the Schottky barrier. The combination of both techniques shows a ten-fold increase in the photocurrent by the acoustic wave. Moreover, the observed spatial redistribution of the current is attributed to the interplay between the contact properties at the source and drain electrodes and the charge carrier dynamics induced by the surface acoustic wave. The ratio of acoustic wavelength to the electrical channel length is found to impact the SAW-induced charge carrier transport. For a channel length shorter than one acoustic wavelength, carriers undergo a seesaw-like motion which changes to charge conveyance for channel lengths comparable or exceeding the acoustic wavelength.

MAIN TEXT

**Introduction**

Since the first successful field-effect magneto-transport measurements on monolayer graphene in 2005[1], 2D materials have attracted widespread research interest. Particularly, semiconducting transition metal dichalcogenides (TMDCs) have shown excellent electronic and optoelectronic properties with bandgaps spanning the ultraviolet to the infrared range[2], strong light matter interaction[3], high charge carrier mobilities[4] and thickness-dependent electrical and optical properties[5,6]. All of these properties of atomically thin layers and their compatibility with the established silicon technology makes them ideal candidates for optoelectronic devices such as light emitting diodes[4,7], photodetectors[7–10], field effect transistors[6,7,11] or photovoltaic solar cells[4,12]. Additionally, the van der Waals nature of their interlayer bond makes their exfoliation and hybrid integration on functional substrates straightforward. For example, LiNbO$_3$ is a strong piezoelectric substrate with applications ranging from optical modulators and waveguide to radiofrequency (RF) filtering using acoustic waves.

In fact, surface acoustic waves (SAWs) are at the foundation of an important class of radio frequency filter devices and have proven to be a versatile tool opening cross-disciplinary applications in nanoscience[13]. When excited on piezoelectric substrates, SAWs provide for instance a contact free method to drive acousto-electric currents (AECs) in low dimensional semiconductor systems[14–17] including 2D materials, in particular semiconducting TMDCs[18–22]. Currently, large efforts are made to advance this approach to a full set of acousto-electric devices employing traditional III-V compound[23–25] or 2D materials[26,27]. Recently, SAWs have been used to manipulate the emission of trions and excitons inside exfoliated $WSe_2$ and $MoSe_2$ monolayers[28–30] and scanning acousto-optoelectric spectroscopy provided a method to faithfully map local defects and trapping sites at room temperature[31]. The strain of standing waves contained in a SAW resonator on GaAs were also used to locally and reversibly tune the optical and structural properties of a $WS_2$ monolayer at room temperature[32]. Concerning optoelectronic devices, SAWs have been shown to improve the properties of $SnS_2$- and $MoS_2$-based photodetectors[33,34].

A ubiquitous challenge encountered in the implementation of large-scale fabrication of TMDC-based optoelectronic devices is the difficulty in forming low resistance Ohmic contacts between metal and semiconducting TMDCs. Here, the metal-semiconductor interface exhibits a complex interplay of Schottky (SB) and tunneling (TB) barriers that has a pronounced impact on electrical transport properties[35–37]. Therefore, the engineering of the metal–semiconductor interfaces, as well as understanding their impacts on the device functionality and physics is of paramount importance for the improvement in optoelectronic device performances.

In this work we investigate the impact of electrical contacts on the performance of an archetypical acousto-electric device. Our device is realized on the strong piezoelectric $LiNbO_3$ material platform and comprises interdigital transducers for SAW generation and an electrically contacted $WSe_2$ multilayer flake. Importantly, this platform offers SAW-induced acousto-electric current (AEC) and spatially resolved photocurrent (PC) measurements which can be combined to fully-fledged acousto-photoelectric current (APEC) spectroscopy on the same device. This combination of methods provides a direct route to characterize the SB and TB formed at the interface between the WSe$_2$ multilayers and the gold electrodes. We show that spatially resolved PC measurements allow for a qualitative investigation of the band bending (BB) induced by the Schottky barriers as well as the spatial extent of their influence on the generated PC. Moreover, the sign of the AEC depends on the SAW propagation direction and confirms the *p*-type conduction of the WSe$_2$ multilayer flakes. The amplitude of the AEC is characterized by the height of the TB formed by the van der Waals gap and the Schottky contact between the gold electrode and the WSe$_2$ flake. The combination of both techniques by measuring the full APEC shows a strong enhancement of the current by the SAW and a change in its spatial distribution. This change is attributed to the interplay between the contact properties at the source and drain and the charge carrier dynamics induced by the SAW. Two exemplary samples are chosen to illustrate the impact of contact asymmetry and the ratio between the source-drain gap and the acoustic wavelength on the acoustoelectric and acousto-photoelectric devices performance. Indeed, Sample 1 with a gap-to-wavelength ratio of 1 shows a current sign dependence on the SAW propagation direction whereas in the case of a 0.5 ratio and high contact resistance asymmetry in Sample 2, no SAW propagation direction dependence is observed.

**Results and Discussion**

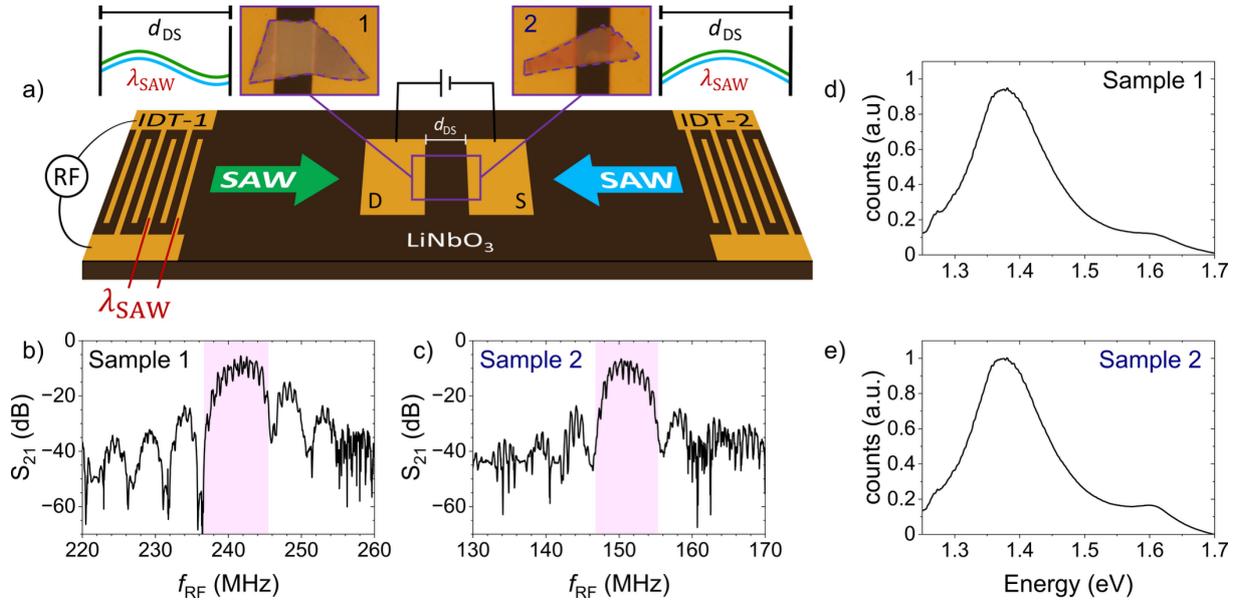

**Figure 1.** Sample layout and pre-characterization – (a) Schematic of the fabricated WSe$_2$-LiNbO$_3$ hybrid samples. A multilayered WSe$_2$ flake is transferred onto two gold electrodes of distance $d_{DS}$ patterned inside a LiNbO$_3$ YZ-cut delay line formed by IDT-1 and IDT-2. The zoom-in highlights the gap-to-wavelength ratios of 1 (Sample 1, left) and 0.5 (Sample 2, right). (b) Scattering parameter $S_{21}$ quantifying the electrical transmission of the SAW across the delay line of Sample 1. SAWs are excited at frequencies ≈ 241 MHz. (c) Scattering parameter S$_{21}$ quantifying the electrical transmission of the SAW across the delay line of Sample 2. SAWs are excited at frequencies ≈ 151 MHz. (d) Representative room-temperature PL spectrum of the emission from the multilayered WSe$_2$ flake between the electrodes in Sample 1. (e) Representative room-temperature PL spectrum of the emission from the multilayered WSe$_2$ flake between the electrodes in Sample 2.

A schematic of the studied WSe$_2$-LiNbO$_3$ hybrid devices is shown in **Figure 1**a. It comprises two interdigital transducers (IDTs) that, by applying a RF voltage, excite forward (green arrow) or backward (blue arrow) propagating SAWs. A mechanically exfoliated WSe$_2$ multilayer is placed on top of two gold source/drain electrodes using a PDMS stamp. The WSe$_2$ circuit is positioned at the center of this SAW delay line. We fabricate several samples to investigate and compare the behavior of different WSe$_2$ multilayers. Two of these samples, referred to as Sample 1 and Sample 2, are discussed in this paper as extreme examples of our model. For completeness, two additional samples with consistent behavior (Sample 3 and Sample 4) are presented in the Supporting Information.

To electrically characterize the functionality of the SAW delay lines, we measure the modulus of the transmission scattering parameter $S_{21}$ as a function of the applied radio frequency $f_{RF}$. The data recorded from Sample 1 and Sample 2 are plotted in **Figure 1**b and **Figure 1**c, respectively, while the complete set of scattering parameters for all samples is provided in the Supporting Information. These data confirm a clear transmission peak at central frequencies of $f_{RF} = 241$ MHz (Sample 1) and 151 MHz (Sample 2) with total bandwidths of $\Delta f_{RF} \approx 8$ MHz marked by the purple boxes. These frequencies correspond to SAW wavelengths $\lambda_{SAW,1} = 14$ μm and $\lambda_{SAW,2} = 23$ μm for Sample 1 and 2, respectively. The distance between the source and drain electrode, is fixed to $d_{DS} = 12$ μm (Sample 1) and 10 μm (Sample 2). Thus, the separations between the electrodes correspond to approximately one acoustic wavelength in the case of Sample 1 and half the acoustic wavelength in the case of Sample 2, as illustrated in the inset of **Figure 1**a. These different geometries enable us to study two different regimes for the transport mechanisms.

The WSe$_2$ flakes are optically characterized by room temperature photoluminescence (PL) spectroscopy in the center between the electrodes. Representative spectra displayed for Sample 1 (**Figure 1**d) and Sample 2 (**Figure 1**e) show a dominant emission band at ≈ 1.4 eV from phonon-assisted recombination for the momentum indirect transition at the fundamental band gap and a weak shoulder at ≈ 1.6 eV due to hot PL from non-thermalized electrons and holes residing at the direct interband transition at the K, K' points at the corner of the 1st Brillouin zone. These emission features are characteristic of few-layered WSe$_2$.[38,39] The less pronounced shoulder of Sample 1 indicates a higher number of layers compared to Sample 2.[38]

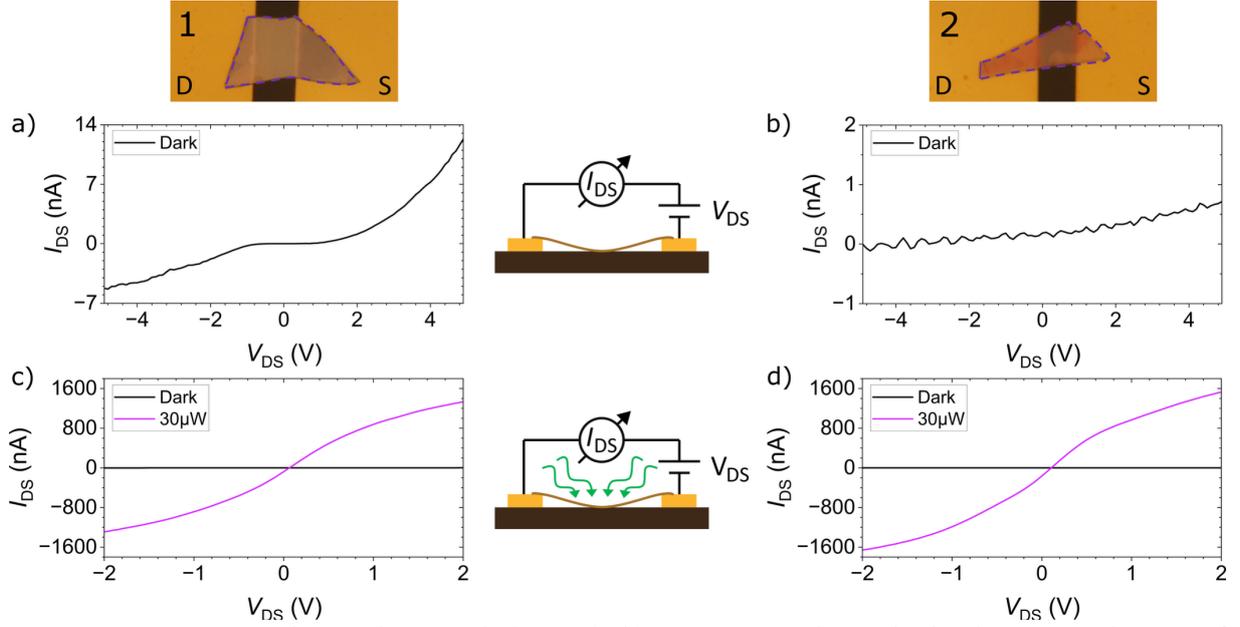

**Figure 2.** Current-voltage characteristics – (a-b) Upper Panel: Optical microscope image of the WSe$_2$ flake in Sample 1 and Sample 2, respectively. Lower Panel: Dark current $I_{DS}$ as a function of the applied voltage $V_{DS}$ for Sample 1 and Sample 2, respectively. (c-d) Current $I_{DS}$ as a function of the applied voltage $V_{DS}$ under global illumination realized by a defocused 30 µW laser illumination on the WSe$_2$ flake for Sample 1 and Sample 2, respectively. The measurement setup is schematically depicted in the two central panels.

We begin by assessing the pristine electrical properties of the TMDC-metal interfaces. We analyze the dark current-voltage ($I_{DS} - V_{DS}$)-characteristics of the WSe$_2$ multilayers which are plotted in **Figure 2**. Without illumination, Sample 1 shows a highly nonlinear, hence almost symmetric dark $IV$-characteristics with values from $I_{DS} = -7$ nA to $+14$ nA for $V_{DS} = -5$ V and $V_{DS} = +5$ V, respectively (**Figure 2**a). Considering the holes as the majority charge carriers in the weakly p-type WSe$_2$, this observation indicates that both, the drain and source, contacts exhibit diode-like IV behavior with similar contact barriers. The minor asymmetry indicates a slightly higher energy barrier at the drain side.

In contrast, the dark $IV$-characteristics of Sample 2, in **Figure 2**b, exhibits a pronounced asymmetry. Here, a measurable current is only observed under positive $V_{DS}$, while $I_{DS}$ under negative bias remains low. These observations imply a high (low) barrier at the drain (source) contact (**Figure 2**b). The low current amplitudes with a maximum of $I_{DS} = +0.7$ nA originate from higher contact barriers compared to Sample 1 *and/or* a lower number of charge carriers due to the smaller thickness of the multilayer.

Under illumination with light energies larger than the direct interband transition, strong absorbance results in the generation of a significant amount of electron and holes that thermalize to the conduction band minimum and valence band maximum and are expected to exhibit rather long lifetimes due to the indirect nature of the fundamental band gap. This photodoping of the full flakes with a defocused green laser ($E_L = 532$ nm $= 2.33$ eV, $P_L = 30$ µW) leads to the *IV*-characteristics of Sample 1 and Sample 2 becoming similar (**Figure 2**c, d). It should be noted that the defocused laser corresponds to a significantly lower power density of $\sim 10^3$, in contrast to the $\sim 10^5$ power density used for the photoluminescence measurements shown in **Figure 1**d and **Figure 1**e. For both samples, the measured current is drastically enhanced, reaching values of $I_{DS} = \pm 1300$ nA (Sample 1) and $I_{DS} = \pm 1600$ nA (Sample 2) at moderate $V_{DS} = \pm 2$ V. The symmetric behavior arises from photogenerated electrons *and* holes contributing to the current. This indicates that for both bias directions at least one charge carrier species is inserted into the contact with the lowest extraction barrier.

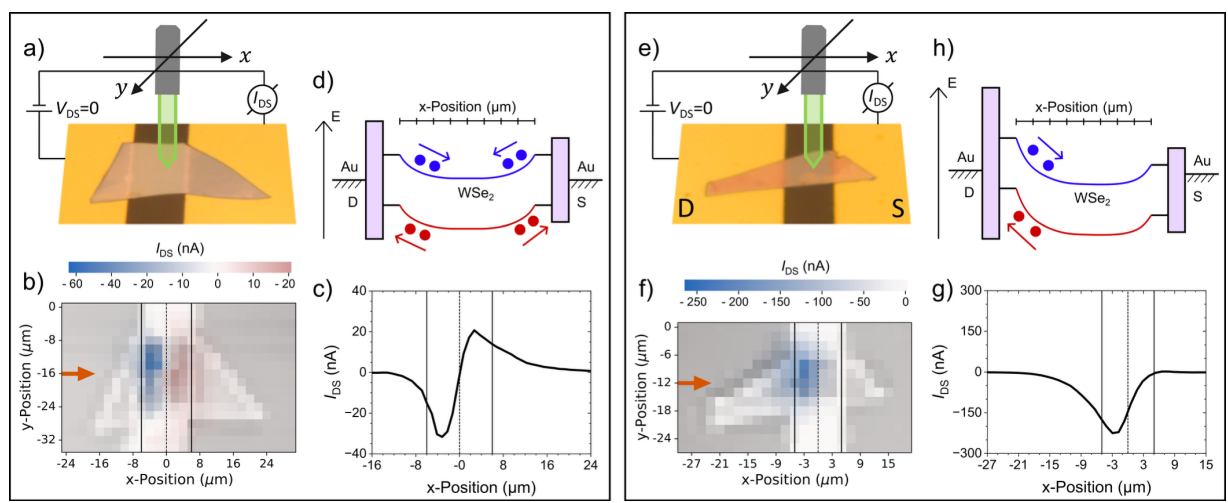

**Figure 3.** Scanning photocurrent measurement – Left panel: Sample 1. (a) Schematical depiction of the scanning PC measurement on the WSe$_2$ flake in Sample 1. (b) False color plot of the PC measured as a function of the laser excitation position. The reflection of the laser is plotted in the background in gray scale. (c) Line plot of the PC measured at position $y = -16$ μm as a function of the laser position x. (d) Schematic of the energy band structure of the WSe$_2$ flake and the contacts at the source and drain electrodes. The work function level of the gold electrodes is followed by purple rectangles indicating the height of the TB created by the van der Waals gap and the Schottky barrier. The bending of the bands (BB) by the SB is shown as well as their influence on the drift of the photogenerated charge carriers shown by the arrows causing the measured current sign. Right panel: Sample 2. (e) Schematical depiction of the scanning PC measurement on the WSe$_2$ flake in Sample 1. (f) False color plot of the PC measured as a function of the laser excitation position. The reflection of the laser is plotted in the background in gray scale. (g) Line plot of the PC measured at position $y = -12$ μm as a function of the laser position x. (h) Schematic of the energy band structure of the WSe$_2$ flake for Sample 2.

Next, we investigate the contact behavior in more details and perform static scanning photocurrent (PC) measurements with no SAW applied. In this case, the laser spot is focused to a spot size of ~1μm and raster scanned across Sample 1 and Sample 2 as shown in the schematics in **Figure 3**a and **Figure 3**e, respectively. At each laser position, we measure the short-circuit PC ($V_{DS} = 0$) generated by the photovoltaic effect[12,40]. Simultaneously, the reflected laser light is detected by a photodiode to image the sample, which we overlay with the PC in **Figure 3**b and in **Figure 3**f. **Figure 3**b shows a PC map of Sample 1 plotted in false-color representation. In this plot, the position of the contact edges and the center of the gap are marked by solid and dashed lines, respectively. These data show a negative PC up to $I_{DS} = -42$nA near the drain contact and a positive PC up to $I_{DS} = +21$nA near the source contact, with similar amplitudes for both polarities. **Figure 3**c, presents a horizontal line scan of these data at a fixed vertical position, $y = -16$ μm (marked by an orange arrow in **Figure 3**b), to further illustrate this behavior. In this plot, the position of the contact edges and the center of the gap is also marked by solid and dashed lines, respectively. This analysis reveals a maximum negative polarity amplitude of $I_{DS} = -31$ nA at the drain electrode and a maximum positive polarity amplitude of $I_{DS} = +21$ nA at the source electrode, respectively. Note that both PC peak positions are located inside the channel at the same distance from the corresponding electrodes as marked by solid lines. Based on previous works on the analysis of the contact between metals and TMDC[41,42], we derived a simplified model of the band structure of the metal-TMDC interface for our devices, which is depicted in **Figure 3**d for Sample 1. In this simplified model, the electrical contact band structure is separated in two main contributions: a band bending (BB) formed by the Schottky barrier (SB) to accommodate the work function difference of the gold electrodes ($\Phi_{Au} = 5.4$ eV) and WSe$_2$ ($\Phi_{WSe_2} = 4.06$ eV),[43] and of a tunneling barrier (TB) (purple rectangle), which takes into account both the tunneling barrier formed by the van der Waals gap between both materials as well as the tunneling effect caused by the SB.[41,42] From the scanning PC measure of Sample 1, we can conclude an almost symmetric spatial profile of the BB, with a slightly dominant BB at the drain electrode leading to a larger measured current $I_{DS}$ at the drain electrode. At the drain contact, the direction of the measured current arises from holes being directly inserted into the drain electrode or by electrons being moved towards the source electrode, as indicated by the red and blue arrows in **Figure 3**d. Both phenomena are induced by the SB forming at the TMDC-metal interface and the corresponding bending of the valence (VB) and conduction (CB) bands, respectively[41,42]. This gives rise to the observed negative current. An analogous reasoning and considering the reversed charge carrier directions applies at the source contract to understand the observed positive PC. In this simplified model, only the TB height is varied to take into account the change in the contact resistance, although both the height and width of the TB affect the charge carrier insertion into the contacts.[44] In a first approximation, the TB at the drain contact is taken to be slightly higher than at the source contact for Sample 1 (see **Figure 3**d). This ansatz is based on the asymmetry in the contact resistance observed in the dark IV-characteristics of the device plotted in **Figure 2**a. However, it should be noted that no information on the TB can be extracted from the scanning PC measurement and further measurements are needed to fully characterize this barrier.

For Sample 2, the false color map (**Figure 3**f) of the scanning PC exhibit a large negative PC up to $I_{DS} = -265$ nA at the drain electrode but no detectable current at the source electrode. These findings are further illustrated in **Figure 3**g which shows the PC at a fixed vertical position, $y = -12$ µm, with a peak of $I_{DS} = -225$ nA at the horizontal position of the drain electrode. Based on our model, depicted in **Figure 3**h, this originates from a large BB at the drain contact and a lower BB at the source. This leads to neglectable charge carrier motions in combination with a high recombination of electrons and holes at the source side. At the drain, electron-hole pairs are efficiently dissociated, with electrons being driven towards the source and holes being directed towards the drain (see **Figure 3**h). Just like for Sample 1, the heights of the TBs of Sample 2 are chosen in a first approximation to be strongly asymmetric, with a high barrier at the drain contact and a much lower one at the source, to reflect the strong bias dependence shown by the sample in its dark *IV*-characteristics (see **Figure 2**b). In consequence, the largest contribution to the measured PC is made by the electrons driven by the BB to the source contact.

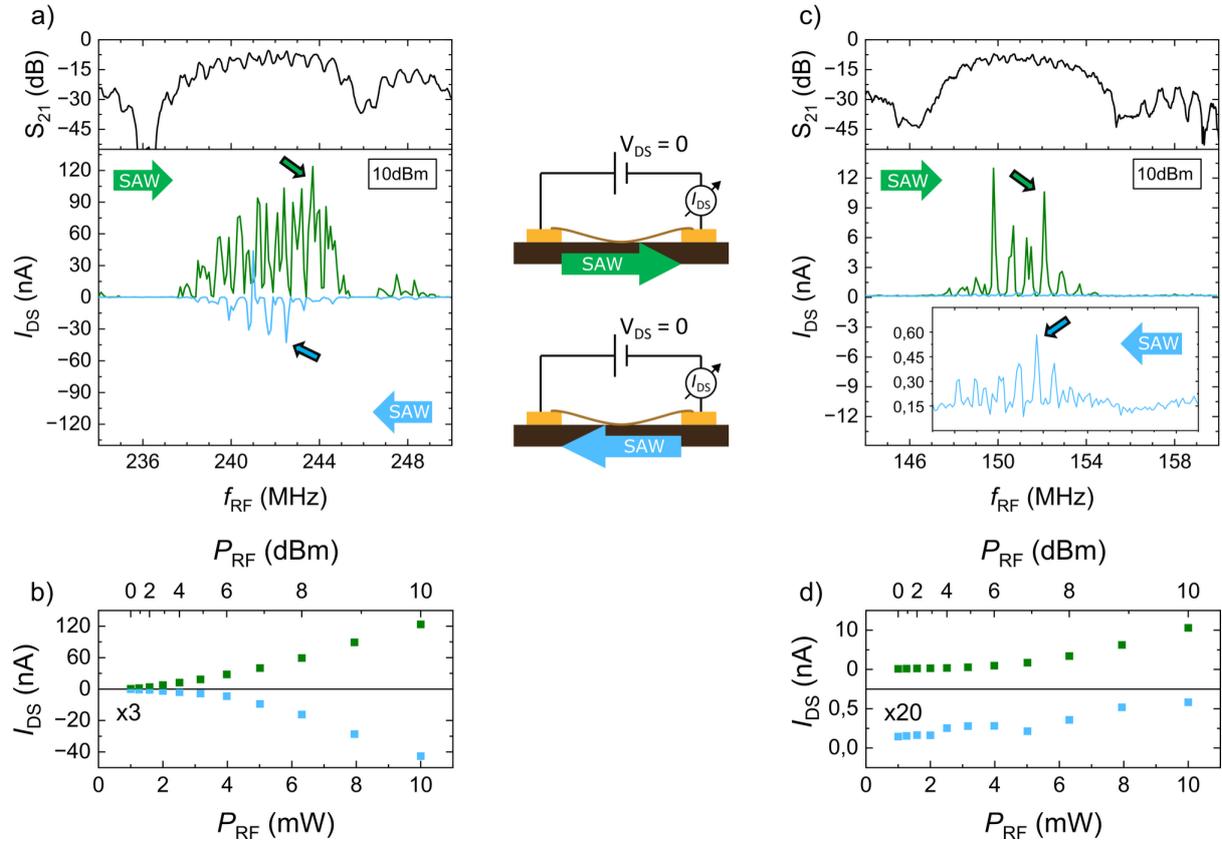

**Figure 4.** Acousto-electric current - (a-c) Upper Panel: Scattering parameter $S_{12}$ of the SAW delay line of Sample 1 and Sample 2, respectively. Lower panel: Acousto-electric current ($I_{DS}$) as a function of the applied SAW frequency for forward (green) and backward (blue) propagating SAWs in Sample 1 and Sample 2, respectively. (b-d) Acousto-electric current ($I_{DS}$) as a function of the applied RF power ($P_{RF}$) for forward (green) and backward (blue) propagating SAWs in Sample 1 and Sample 2, respectively. Center Panel: schematic depiction of the measurement setup.

After these static investigations of the devices' contact behaviors, we study the impact of the TBs and BBs on the charge carrier dynamics driven by the electric field of an applied SAW. In **Figure 4**, we compare the dependence of the induced short circuit ($V_{DS} = 0$) acousto-electric current ($I_{DS}$) on the SAW propagation direction and power. For both SAW propagation directions, the acousto-electric current (AEC) in Sample 1 is exclusively observed within the SAW's $S_{21}$-transmission band, plotted in the upper panel of **Figure 4**a. This provides direct evidence that the effect is indeed driven by the SAW. At a constant applied power of $P_{RF} = 10$ dBm (**Figure 4**a), Sample 1 shows clearly the expected SAW directionality with a positive AEC of up to $I_{DS} = +125$ nA for a forward propagating SAW (green) and negative AEC below $I_{DS} = -43$ nA for a backward propagating SAW (blue). The observed polarities of $I_{DS}$ for forward and backward propagating SAWs confirm that the detected current arises from *p*-type charge carriers as expected for WSe$_2$. Moreover, the difference of the AEC magnitude between both propagation directions is in full agreement with the TBs schematically depicted in **Figure 3**c, underpinning the more effective insertion of holes into the source electrode and confirming our first evaluation of the TBs based on the dark IV-characteristics measured previously (**Figure 2**a). Finally, we evaluate the $P_{Rf}$-dependance of the AEC (**Figure 4**b) at constant SAW frequencies of $f_{SAW} = 243.7$ MHz and $f_{SAW} = 242.5$ MHz (highlighted by arrows in **Figure 4**a) for forward (green symbols, upper panel) and backward propagating (blue symbols, lower panel) SAWs. These data show monotonous linear dependencies expected for a regular AEC for *p*-type conduction[15].

Analogous data for Sample 2 is shown in **Figure 4**c. Again, the AEC is only observed within the SAW's $S_{21}$-transmission band, plotted in the upper panel. In contrast to Sample 1, these data don't exhibit the anticipated polarity of the AEC with a maximum magnitude of $I_{DS} = +13$ nA at $P_{RF} = 10$ dBm exclusively for the forward propagating SAW (green line, upper panel). In strong contrast, the backward propagating SAW (blue line, lower panel) induces only a weak and positive AEC with a maximum of $I_{DS} = 0.6$ nA. This anomalous behavior is confirmed in $P_{RF}$-dependent measurements at $f_{SAW} = 152.1$ MHz and $f_{SAW} = 151.7$ MHz (**Figure 4**d). Again, these data show the expected linear dependence exclusively for a forward propagating SAW and $P_{RF} > 7$ dBm. For a backward propagating SAW, a positive AEC is detected which remain weak compared to the forward propagating SAW. The reason underlying the observed anomalous AEC for the backward propagating SAW is the extraction of holes from the WSe$_2$ semiconductor into the source electrode. This motion, facilitated by the SAWs electric field, is in opposite direction to that of the SAW. Vice versa, extraction at the drain electrode is suppressed by the strongly dominating TB at the drain side (see **Figure 3**g).

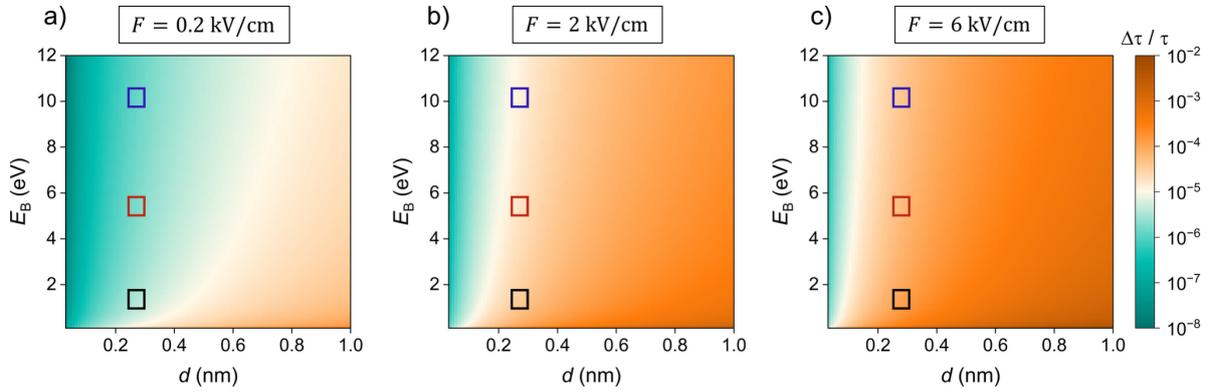

**Figure 5.** Tunneling probability – False color plot of the modulation of the tunneling time $\Delta\tau/\tau$ through the vdW and Schottky tunnel barrier as a function of the barrier energy $E_B$ and width $d$ for three in-plane SAW-induced electric field amplitudes: (a) 0.2, (b) 2 and (c) 6 kV/cm. Representative tunneling barrier heights are indicated by the work function of gold $\Phi_{Au}$ (red box), the gold–WSe$_2$ work function difference $\Phi_{Au} - \Phi_{WSe_2}$, and the effective barrier height $E_B$ (black box).[42,43]

In the next step, we investigate the modulation of the tunnelling barriers and the resulting tunneling rates by the dynamic electric field of the SAW. This SAW-driven mechanism has been identified to govern carrier transfer within quantum confined systems in nanowires.[45] To this end, we calculate the tunneling rate through a rectangular barrier as a rough estimation of the vdW and Schottky barriers using a Wentzel-Kramers-Brillouin (WKB) model for different values of in-plane component of SAW's electric field, $F_\parallel$. The tunneling rate of electrons and holes as a function of $F_\parallel$ is dependent on the height $E_B$ and width $d$ of the rectangular barrier and is given by[46]

$$\tau^{-1}_{\text{tunnel,h}} = \exp[-\frac{4\sqrt{2m_h^* E_B^3}}{3\hbar e F_\parallel}]\left(1 - \left(1 - \frac{e \cdot F_\parallel \cdot d}{E_B}\right)^{\frac{3}{2}}\right) \quad (1)$$

with $m_e^* = 0.36 m_0$ and $m_h^* = 0.30 m_0$ denoting representative values for the effective masses of electrons and holes in monolayer WSe$_2$.[47] Since $m_e^* \approx m_h^*$, we here exemplarily discuss the results for holes. The dimension-less modulation of the tunneling time $\Delta\tau/\tau$ driven by the switching between $\pm F_{\parallel,\text{max}}$ over one acoustic period is quantified by[45]

$$\Delta\tau/\tau = \frac{|\tau_{\text{tunnel}}(-F_{\parallel,\text{Max}}) - \tau_{\text{tunnel}}(+F_{\parallel,\text{Max}})|}{\tau_{\text{tunnel}}(F_\parallel = 0)}. \quad (2)$$

In **Figure 5**, we plot the $\Delta\tau/\tau$ dependency on $d$ and $E_B$ for $F_{\parallel,\text{Max}} = 0.2$ kV/cm (a), $F_{\parallel,\text{Max}} = 2$ kV/cm (b), and $F_{\parallel,\text{Max}} = 6$ kV/cm (c) in false color representation. Note that the highest accessible electrical field $F_{\parallel,\text{Max}} = 6$ kV/cm is reached in our devices for an applied RF power of $P_{\text{RF}} = 21$ dBm[31], which is significantly higher than the RF power applied in our experiment. The color boxes mark the most relevant combinations of $d$ and $E_B$. In the following analysis, we assume $d = 0.27$ nm and consider three representative values for $E_B$. The chosen $d$ was obtained by Kang and coworkers in a computational study of gold contacts to WSe$_2$ monolayers.[42] In the same work, the effective barrier height $E_B = 10.2$ eV was determined (blue box). The other representative values of $E_B$ are the work function of gold $E_B = \Phi_{\text{Au}} = 5.4$ eV (red box) and the difference between the work functions of gold and WSe$_2$, $E_B = \Phi_{\text{Au}} - \Phi_{\text{WSe}_2} = 1.34$ eV[43] (black box). The latter represents the lower bound of the Schottky barrier height without any vdW tunneling barrier and the former the height of the tunnel barrier without the addition of the Schottky barrier. Our model confirms the expected trend for all cases shown in **Figure 5**: the modulation $\Delta\tau/\tau$, given by equation (2), increases with increasing $F_{\parallel,\text{Max}}$ and is most pronounced at low $E_B$ and high $d$. However, the maximum modulation for all studied electrical fields and for a significantly expanded range of values for $E_B$ and $d$ remains below $\leq 10^{-2}$. Thus, we can clearly conclude that the dynamic modulation of the tunnel rate by the SAW does not contribute significantly to the observed changes of the electrical current under SAW excitation. Additionally, the non-linearity in the $P_{rf}$-dependency of this effect is in direct disagreement with our observation in **Figure 4**b and d. Therefore, all effects observed in our experiments predominantly result from SAW-driven charge carrier dynamics inside the WSe$_2$ flake and for all further considerations, the effect of barriers can be considered as static.

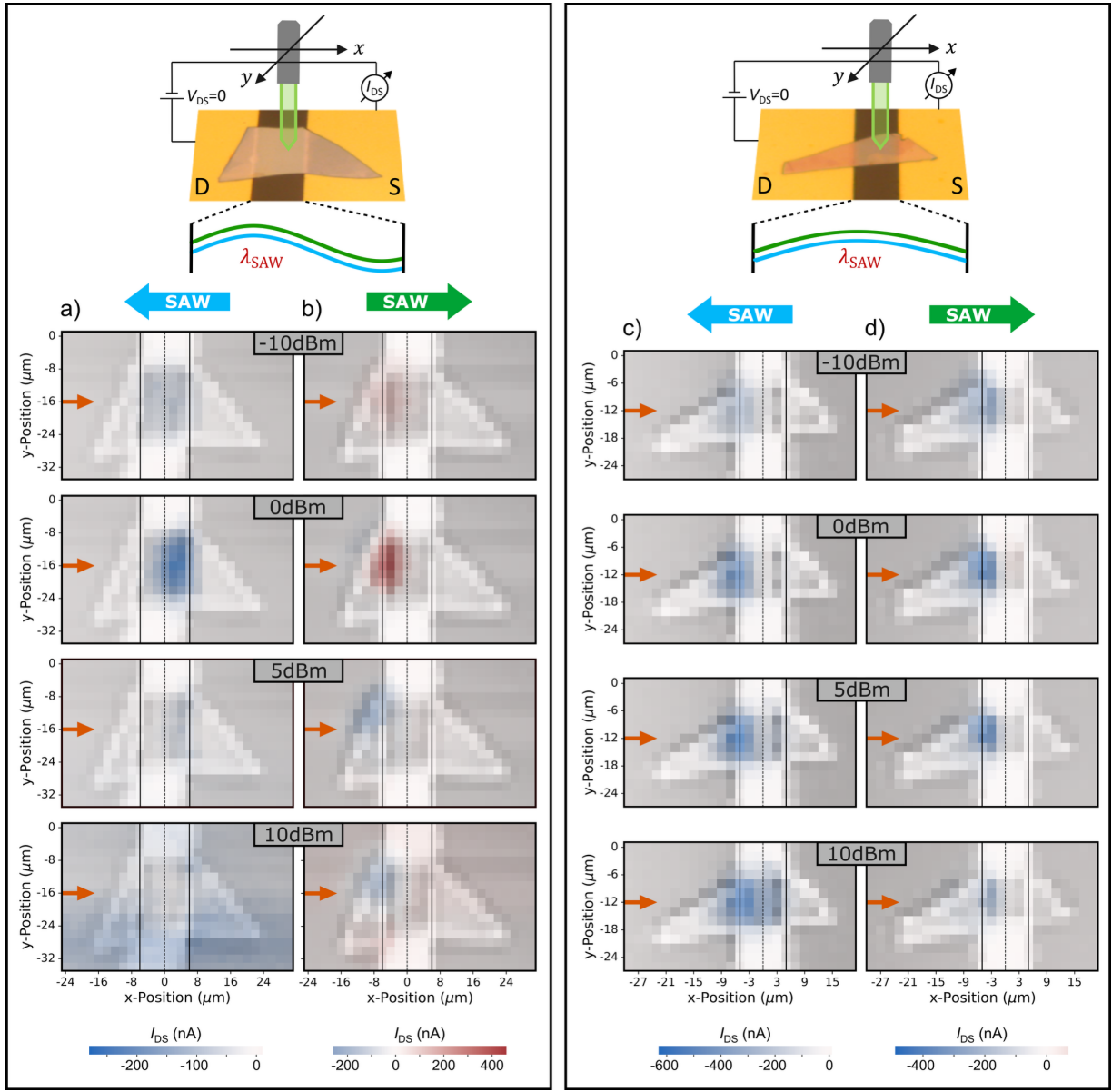

**Figure 6.** Scanning acousto-photoelectric current measurement – (a-c) False color plot of the photocurrent as a function of the laser excitation position $(x, y)$ for a backwards propagating SAW for an applied $P_{RF}$ = -10, 0, 5 and 10 dBm for Sample 1 (a) and 2 (c), respectively. The color scale of the measurement series is placed in the last panel. (b-d) False color plot of the photocurrent as a function of the laser excitation position $(x, y)$ for a forward propagating SAW for an applied $P_{RF}$ = -10, 0, 5 and 10 dBm for Sample 1 (b) and 2 (d), respectively. The color scale of the measurement series is placed in the last panel. The schematics in the top panel illustrates the gap-to-wavelength ratios of Sample 1 and Sample 2.

To investigate the direct impact of the TBs and BB in acousto-electric devices, we combine scanning PC spectroscopy with SAW. **Figure 6** presents spatial scans of the short-circuit acousto-photoelectric current (APEC) in false color representation for Sample 1 (a,b) and Sample 2 (c,d) for backward (a,c) and forward propagating SAWs (b,d). Schematics show the measurement geometry, the corresponding gap-to-wavelength ratios and the propagation directions of the applied SAWs. Moreover, the applied RF power is tuned from $P_{RF}$ = -10 to 10 dBm from the top to bottom panels. The data for a backward propagating SAW applied on Sample 1 (**Figure 6**a) shows two striking effects. First, the polarity of the current at the drain electrodes changes from a negative PC (cf. **Figure 3**b) to a positive APEC. Second, we observe a strong increase of the negative APEC amplitude at the source contact when $P_{RF}$ increases from $-10$ dBm to 0 dBm. At these power levels, the current at the drain electrode remains low. When photo-generating carriers in the vicinity of the source electrode, we observe a negative current $I_{DS} = -280$ nA at $P_{RF} = 0$ dBm. The negative polarity of the APEC is consistent with a unipolar acousto-electric transport of photogenerated holes from the source to the drain. The measured currents corresponds to a relative enhancement of the unperturbed PC (cf. **Figure 3**b) of $|I_{DS}(APEC)/I_{DS}(PC)| \sim 7$ and $|I_{DS}(APEC)/\max(I_{DS}(AEC), I_{DS}(noise))| \sim 1400$ of the dark AEC, if larger than the noise level of the experiment, at $P_{RF} = 0$ dBm. Most interestingly, increasing the SAW amplitude at $P_{RF} = 5$ dBm, quenches the large negative current induced by the photogenerated carriers close to the source electrode. This suppression can be attributed to bipolar acousto-electric, i.e.

transport of both electrons and holes from source to drain by the SAW. Thus, no net charge is conveyed by the SAW and, therefore, the acousto-electric effect does not add a net contribution to the total current. When increasing the RF power to $P_{\text{RF}} = 10$ dBm, the total current is fully dominated by AEC of the intrinsic *p*-type carriers, which are also present without photoexcitation (cf. **Figure 4**). Thus, our spatial scan shows a position-independent homogenous negative current.

For forward propagating SAW, the situation reverses: at low $P_{\text{RF}} = -10$ dBm and $P_{\text{RF}} = 0$ dBm, unipolar transport of holes is induced from the drain to the source electrode. This leads to a positive APEC reaching up to $I_{\text{DS}} = +460$ nA when carriers are photogenerated in vicinity of the drain contact. Comparison to the unperturbed PC (cf. **Figure 3**) and dark AEC (cf. **Figure 4**) yields enhancements of $|I_{\text{DS}}(APEC)/I_{\text{DS}}(PC)| \approx 11$ and $|I_{\text{DS}}(APEC)/\max(I_{\text{DS}}(AEC), I_{\text{DS}}(noise))| \approx 350$, respectively. Thus, both the direction and the peak position of the APEC is reversed compared to the backward propagating SAW. Moreover, and fully analogous to the case of the backward propagating SAW, at moderate ($P_{\text{RF}} = 5$ dBm) and high ($P_{RF} = 10$ dBm) SAW amplitudes, the transition from unipolar to bipolar acousto-electric transport occurs with a positive net AEC under strong SAW-modulation.

In contrast, analogous data of Sample 2 presented in **Figure 6**c and **Figure 6**d reveal different behavior compared to that of Sample 1. First, we observe negative current close to the drain electrode for both SAW propagation directions and when increasing $P_{\text{RF}}$. This observation mirrors the scanning PC data in **Figure 3**f and **Figure 3**g. Second, we do not observe a clear transition from the unipolar to bipolar acousto-electric transport. Analogous to Sample 1, we find a clear enhancement of $I_{\text{DS}}$. From our data, we extract for a backward propagating SAWs $I_{\text{DS}} = -540$nA (cf. **Figure 6**c) corresponding to relative enhancements $|I_{\text{DS}}(APEC)|/|I_{\text{DS}}(PC)| \approx 2$ and $|I_{\text{DS}}(APEC)|/\max(I_{\text{DS}}(AEC), I_{\text{DS}}(noise)) \approx 2700$, respectively. For a forward propagating SAW, we find $I_{\text{DS}} = -480$nA (cf. **Figure 6**d) and enhancements $|I_{DS}(APEC)|/|I_{DS}(PC)| \approx 2$ and $|I_{DS}(APEC)|/\max(I_{DS}(AEC), I_{DS}(noise)) \approx 2400$.

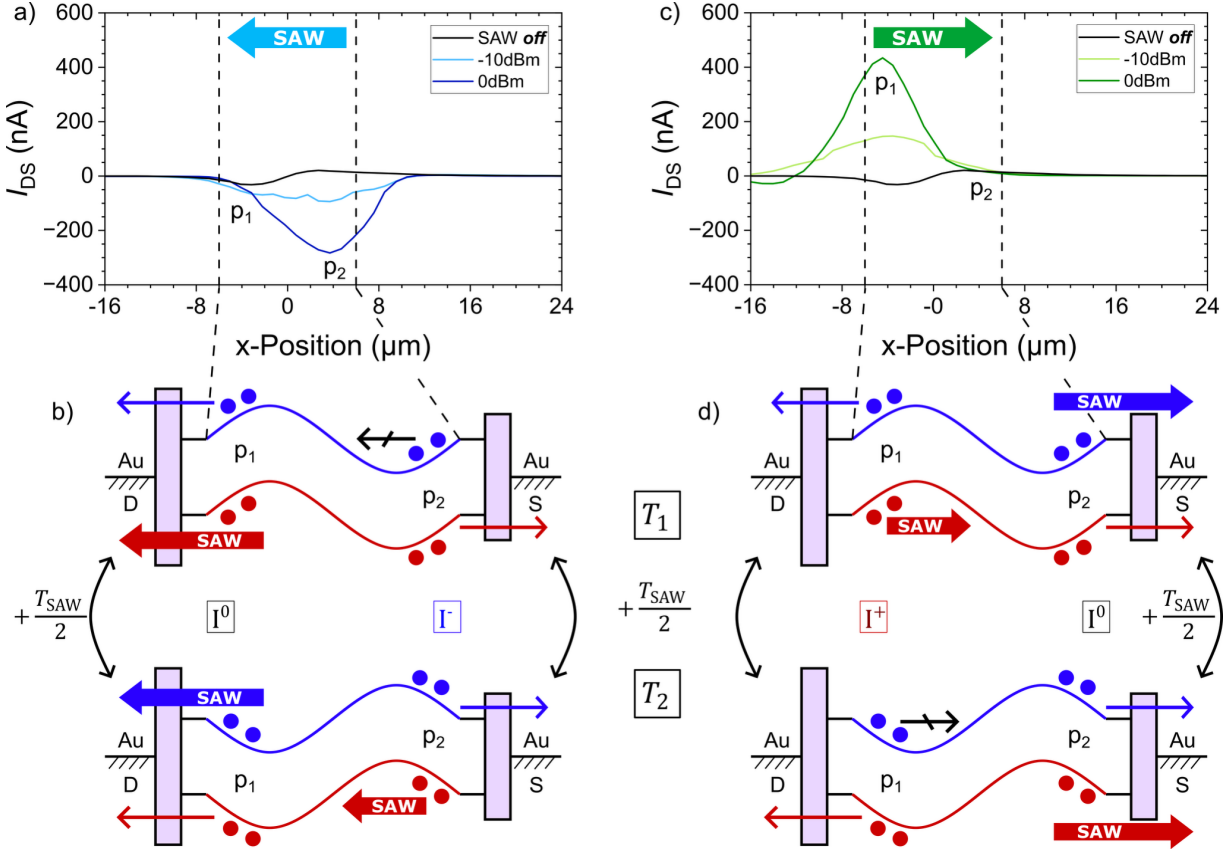

**Figure 7.** Charge carrier dynamics in Sample 1 – (a-c) Line plot of the PC measured at position $y = -16$ μm (orange arrows in **Figure 6**a-b) as a function of the laser position x with and without a backwards (a) and forward (c) propagating SAW. (b-d) Schematics of the band structure model under the modulation of the SAW electric field at two different times, $T_1$ and $T_2$, during the SAW period. The motion of the charge carriers induced by the backward (b) and forward (d) propagating SAW is illustrated by arrows.

In the following, we analyze in detail the APEC of Sample 1 and Sample 2 in the unipolar transport regime under SAW modulation. We identify the main contributing charge carrier transport and extraction processes and their time-dependence during one acoustic cycle. We begin our analysis with Sample 1 and plot the $x$-position dependent APEC ($y = -16$ μm, marked by orange arrows in **Figure 6**a and **Figure 6**b) in **Figure 7**a (backward propagating SAW) and **Figure 7**c (forward propagating SAW). The backward propagating SAW (**Figure** 7a) exhibits the previously discussed negative peak close to the source electrode (position p₂ in the diagram) at $P_{RF} = -10$dBm (light blue) and $P_{RF} = 0$dBm (dark blue) and zero current at the drain electrode (position p₁). For comparison, we additionally plot the sole photocurrent without SAW (black).

In **Figure 7**b, we combine the band diagrams of the tunnel barriers of **Figure 3**c with the band modulation induced by the backward propagating SAW of Sample 1. For simplicity, the band bending induced by the Schottky barrier is not depicted here as the SAW-induced potential of a few volts[31] is commensurate with the difference between the work functions of gold and WSe₂.[43] As introduced at the beginning, the SAW wavelength in this sample is $\lambda_{SAW,1} = 14$ µm and the distance between the source and drain electrodes, depicted in **Figure 1**a, is fixed to $d_{DS} = 12$ µm. Thus, the separations between the electrodes correspond to approximately one acoustic wavelength as shown schematically in **Figure 1**a and **Figure 6**. This allows us to identify the main transport and extraction processes contributing to the measured APEC at two designated times $T_1$ and $T_2 = T_1 + T_{SAW}/2$ during the acoustic cycle. At position p₁, we see that the SAW's electric field drives electrons toward the drain at time $T_1$ (light blue arrow) and the analogous process occurs for holes at $T_2$ (light red arrow). These electric field induced directional extraction at the drain electrode are the opposite direction to the SAW propagation and lead to a no net current when averaged over one acoustic cycle. Additionally at position p₁, the holes are directly transported by the SAW at the VB maximum towards the drain electrode at time $T_1$ (bold red arrow). The analogous process occurs then for electrons at the CB minimum half a period later at $T_2$ (bold blue arrow). Again, these two transport processes also lead to a zero net current when averaged over a full acoustic cycle. In sum, no net current is generated for the photogeneration of free charge carriers at the drain electrode, as shown in **Figure 7**a at position p₁.

The electric field induced extraction of holes and electrons in the opposite direction of the propagating SAW is also present at position p₂ close to the source electrode. Here, the holes are injected in the source electrode at time $T_1$ (light red arrow) and the same process occurs for the electrons at time $T_2$ (light blue arrow), leading in average to no net current. However, we observe a dominant transport of holes at the VB maximum propagating with the SAW towards the drain electrode (time $T_2$, bold red arrow) compared to electrons being transported at the CB minimum (time $T_1$, crossed-out black arrow). This leads to the negative APEC measured at p$_2$ in **Figure 7**a. The less effective transport of electrons compared to holes over longer distances might be caused by increased recombination with a higher number of majority holes (*p*-type material). In addition, electrons have been reported to exhibit lower mobilities than holes in WSe$_2$.[48,49] Indeed, using $\mu_e = 30\,\frac{cm^2}{Vs}$ and $\mu_h = 180\,\frac{cm^2}{Vs}$ obtained as lower boundaries by Allain and coworkers for single layer WSe$_2$[49], we calculate for the drift velocities of electrons and holes, given by

$$v_e = \mu_{e,h} \cdot F_{SAW,} \quad (3)$$

$v_e = 600\,\frac{m}{s}$ and $v_h = 3600\,\frac{m}{s}$ for a moderate SAWs electrical field of $F_{SAW} = 2\,kV/cm$, confirming that only the drift velocity of holes is comparable to the phase velocity of the SAW, which is the more likely cause for the weak transport of electron by the SAW compared to the holes. This difference in the transport of electrons and holes by the SAW leads to the overall negative APEC measured at the source electrode (position p$_2$) at moderate $P_{rf}$.

For the forward propagating SAW plotted in **Figure 7**c, the positive APEC at the source side (position p$_1$) and a zero net current at the drain (position p$_2$) can be explained analogously. Here, the model of the barriers combined with the band modulations and all corresponding direct transport process are considered for the reversed SAW propagation direction in **Figure 7**d. At the source electrode (position p$_2$), the SAW-induced transport and the electric field driven injection of electrons, represented by the bold blue arrow at time $T_1$ and the light blue arrow at time $T_2$ respectively, are balanced by the analogous processes for the holes, represented by the bold red and light red arrow respectively. This leads to the measured zero net current at the source electrode. At the drain electrode (position $p_1$) however, the dominating transport of holes (bold red arrow at time $T_1$) leads to the measured strong positive APEC. Finally, the higher net current measured for the forward propagating SAW is in agreement with the previously assumed lower tunnel barrier at the source electrode obtained from the electrical characterization (**Figure 2**a) and the acousto-electric current measurement (**Figure 4**a-b).

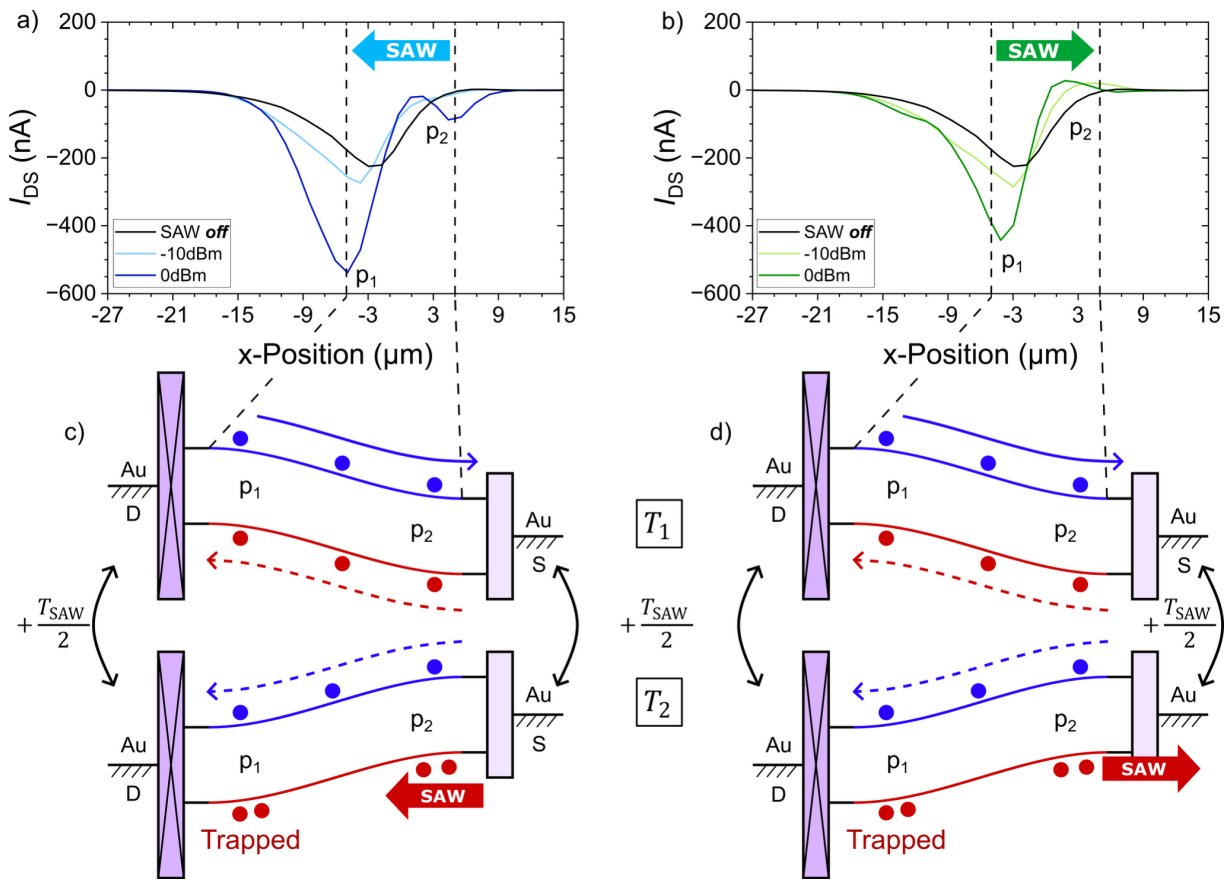

**Figure 8**. Charge carrier dynamics in Sample 2 – (a-c) Line plot of the PC measured at position $y = -12$ µm (orange arrow in **Figure 6**c-d) as a function of the laser position x with and without a backwards (a) and forward (c) propagating SAW. (b-d) Schematics of the band structure under the modulation of the SAW electric field at two different times, $T_1$ and $T_2$, during the SAW period. The motion of the charge carriers induced by the backward (b) and forward (d) propagating SAW is illustrated by arrows.

We continue our analysis with Sample 2 and we plot the $x$-position dependent APEC ($y = -12$ μm, marked by orange arrows in **Figure 6**c and **Figure 6**d) in **Figure 8**a (backward propagating SAW) and **Figure 8**b (forward propagating SAW), respectively. For the backward propagating SAW, we clearly observe a negative current peak at position p$_1$ (drain electrode) for $P_{RF} = -10$ dBm and $P_{RF} = 0$ dBm with a strong enhancement compared to the sole PC (black line). Strikingly, as previously shown in **Figure 6**d, we do not observe the expected change of the current polarity dependent on the SAW direction at position p$_1$. The measured APEC remains negative, and its amplitude is enhanced by the SAW compared to the unmodulated PC (black line) for the forward propagating SAW. In contrast, at the position p$_2$ (source electrode), a weak negative current is observed for the backward propagating SAW for $P_{RF} = 0$ dBm and a weak positive current appears for the forward propagating SAW for $P_{RF} = -10$ dBm and $P_{RF} = 0$ dBm. These positive currents are the only indication of a small directional dependence. Additionally, the negative peak position travels towards the drain electrode, from $x \approx -1$μm to $x \approx -5$μm, as the applied $P_{rf}$ increases, both for the forward and backward propagating SAW.

These observations made for Sample 2 are in strong contrast to those for Sample. 1. These dissimilarities arise from the difference in the contacts barriers (cf. **Figure 3**h) and from the difference in the sample geometries. In Sample 2, the SAW wavelength is $\lambda_{SAW,2} = 23$ μm for a frequency $f_{RF} = 151$ MHz and the distance between the source and drain electrodes, depicted in **Figure 1**a and **Figure 6**, is fixed to $d_{DS} = 10$ μm. The separation between the electrodes thus corresponds to approximately half of an acoustic wavelength as shown schematically in **Figure 3**h. This wavelength to gap ratio of approximately 0.5 significantly alters the underlying charge carrier dynamics in the device.

Combining the model of the TBs from **Figure 3**h with the band edge modulation induced by the backward propagating SAW in **Figure 8**c, the response of the charge carriers on the SAW's electric field can be considered as a seesaw-like motion. This contrasts with fully-fledge transport by the full SAW modulation present in Sample 1. At position p$_1$, the photogenerated electrons are directed towards the source electrode by the SAW-induced bending of the CB at time $T_1$ (light blue arrow). The equivalent motion of electrons towards the drain electrode half an acoustic cycle later at time $T_2$ (dashed light blue arrow) does however not contribute to the net current due to the high drain TB obtained from the electrical characterization (**Figure 2**b) and the acousto-electric current measurement (**Figure 4**c-d). This suppresses any charge carrier injection into the drain electrode and, consequently, the injection of holes by the SAW into the drain at time $T_1$ (dashed light red arrow) is blocked. Additionally, the strong band bending present at the drain (see **Figure 3**h) induces an electric field which counteracts their motion towards the source at time $T_2$ (red circles tagged "Trapped"). These two processes lead to an accumulation of the holes at the drain, causing a screening of the electric field and a change in the BB at position p$_1$. This effect is assumed to be the cause of the negative APEC peak motion toward the drain electrode in **Figure 8**a. When averaging all processes over a full acoustic cycle, only the electron motion contributes to a net current and leads to the negative APEC measured at position p$_1$ for the backward propagation SAW (cf. **Figure 8**a).

Looking now at the position p$_2$ in **Figure 8**c diagram, photogenerated holes are expected to lead to a zero net current as they balance the motion of the electrons at both time $T_1$ and $T_2$ during the acoustic cycle. However, at time $T_2$, the holes are generated at a stable maximum of the VB and, subsequentially, are transported by the SAW towards the drain electrode due to their high mobility (bold red arrow). This effect leads to an increased relative number of electrons being extracted at the source and results in the weak negative current measured at position p$_2$ for the backward propagating SAW at $P_{RF} = 0$ dBm.

In the case of the forward propagating SAW, the negative current measured at position p$_1$ (cf. **Figure 8**b) is caused by the same type of SAW-induced charge carrier dynamics as for the case of the backwards propagating. Here, the combination of the model from **Figure 3**h with the band modulation induced by the SAW in **Figure 8**d shows how identical both instances are. Thus, the "trapped" holes due to the high TB and BB at the drain electrode combined with the seesaw-like electron motion induced by the SAW leads to the direction independent negative APEC measured for Sample 2 as well as the same peak position change as for Sample 1. In contrast, the positive current measured at position $p_2$ is caused by the transport of the holes towards the source electrode by the SAW (bold red arrow). Therefore, the dominant contribution to current at this position is transport of the holes by the SAW. This contribution also gives rise to the, compared to Sample 1, weak directional dependence in Sample 2.

**Conclusion**

In conclusion, we investigated the impact of contact barriers on the electrical transport in SAW-based TMDC-LiNbO$_3$ acousto-electric devices employing a combination of acoustoelectric and scanning photocurrent techniques. Furthermore, we developed a model which consistently describes the impact of the contact resistance and contact distance on the charge carrier dynamics induced by the SAW in acousto-photoelectric devices. Two samples were compared: Sample 1 had symmetric contacts and a wavelength-to-gap ratio of 1, while Sample 2 had asymmetric contacts and a ratio of 0.5. Sample 1 showed direction-dependent photocurrent and strong bipolar transport behavior, indicating effective SAW-driven charge carrier transport. In contrast, Sample 2 exhibited unidirectional, electron-dominated current largely independent of SAW direction, due to high contact asymmetry and a reduced gap ratio. These results highlight contact resistance and SAW wavelength-to-gap ratio as key design parameters in optimizing acousto-electric and acousto-photoelectric device performance.

The impact of the former was expected and contact engineering for 2D materials has become a strong part of the research on 2D material-based electronic and optoelectronic devices.[35–37] Several approaches and techniques are being investigated to solve this issue:

work function engineering to reduce the Schottky barrier height[35], chemical doping[35], ultra-high vacuum metal deposition[35], metal mediated exfoliation[36], van der Waals metal as contact material[37], and so on, each with their specific advantages and limitations. All in all, a systematic characterization of the contact behavior is necessary when designing TMDC-based electronic and optoelectronic devices, given the considerable sample-to-sample variability observed even under similar fabrication conditions. Concerning the acoustic wavelength-to-gap ratio, our results show that achieving a source-drain separation larger than the acoustic wavelength is key to enabling bipolar transport. This design parameter, which was previously not fully considered, is crucial to fully take advantage of the SAW direction dependent acousto-electric current for the effective design of microscale acousto-electric and acousto-photoelectric devices. Finally, the studied device concept can be directly used for fully-fledged acousto-optoelectric spectroscopy[31] opening routes to probe coupled charge carrier dynamics controlled by static and dynamic electric fields. This method has been established for multi-channel studies[50] and correlation spectroscopy[51] rendering it particularly attractive to probe SAW-driven charge transfer processes on electrically contacted van der Waals systems with voltage-controlled interlayer excitons.[52–54]

**Supporting Information**

Full set of reflection coefficients and experimental studies for additional samples.

**Methods**

*Sample Fabrication and Characterization:* Ti/Au (10 nm, 50 nm) IDTs and contact electrodes are fabricated on Y-cut LiNbO$_3$ using an optical lithography and lift-off process. IDTs are aligned in a Z-propagating delay line geometry with a total length of 5000 µm. The IDT designs of Sample 1 and Sample 2 facilitate generation of $f_{SAW} = 241$ MHz ($\lambda_{SAW} = 14.5$ µm) and $f_{SAW} = 151$ MHz ($\lambda_{SAW} = 23.1$ µm), respectively. The RF scattering parameters of the delay lines is determined using a vector network analyzer (Pico Technology PicoVNA 106). Multilayer WSe$_2$ flakes are mechanically exfoliated from a bulk crystal and subsequentially transferred onto the contact electrodes using a dry transfer technique employing PDMS stamps. PL spectroscopy is performed at room temperature using a modular confocal Photoluminescence / Raman microscopy system (Oxford WItec alpha300 R) with a $\lambda = 488$nm laser.

*Scanning photocurrent and acousto-electric measurements*: All experiments were conducted at room temperature under ambient conditions. A 532 nm diode-pumped solid-state laser (time-averaged power 30 µW in cw mode) was focused to a diffraction-limited spot (diameter 1µm) via a 50x objective (NA 0.60). The sample was scanned with a motorized DC stage, while the photocurrent was simultaneously recorded in a short-circuit configuration ($V_{\text{DS}} = 0$) using a source measure unit (Keithley 2450) connected to the two gold electrodes. The reflected laser signal was quantified using a photodiode (Thorlabs DET025AL). For acousto-electric measurements, SAWs were excited using a standard RF signal generator (Stanford Instruments SG382) connected to the IDT.

ASSOCIATED CONTENT

The following files are available free of charge.

AUTHOR INFORMATION

**Corresponding Author**

*emeline.nysten@uni-muenster.de

**Author Contributions**

The manuscript was written through contributions of all authors. All authors have given approval to the final version of the manuscript. ‡These authors contributed equally. (match statement to author names with a symbol)

**Funding Sources**

Any funds used to support the research of the manuscript should be placed here (per journal style).

**Notes**

Any additional relevant notes should be placed here.

ACKNOWLEDGMENT

This work is funded by the Deutsche Forschungsgemeinschaft (DFG, German Research Foundation) in the framework of SPP2244 and individual research grants E.D.S.N. 701979; U.W. WU637/4-2, 7-1; and individual research grant H.J.K. 465136867 and 494822818.

ABBREVIATIONS

| | |
|---|---|
| SAW | Surface acoustic wave |
| TMDC | Transition metal dicalchogenide |
| RF | Radiofrequency |
| PL | Photoluminescence |
| AEC | Acousto-electric current |
| PC | Photocurrent |
| APEC | Acousto-photoelectric current |
| SB | Schottky barrier |
| TB | Tunneling barrier |
| BB | Band bending |
| VB | Valence band |
| CB | Conduction band |

# Impact of Electrical Contacts on Transition Metal Dichalcogenides-Based Acoustoelectric and Acousto-Photoelectric Devices

*Benjamin Mayer[1], Felix M. Ehring[1], Clemens Strobl[1], Matthias Weiß[1], Ursula Wurstbauer[1,2], Hubert J. Krenner[1] and Emeline D. S. Nysten[1,]\**

[1]Physikalisches Institut, Universität Münster, Wilhelm-Klemm-Straße 10, 48149 Münster, Germany

[2]Center for Soft Nanoscience (SoN), Universität Münster, Busso-Peus-Str.10, 48149 Münster

emeline.nysten@uni-muenster.de

## 1. Reflection coefficients of Sample 1-4

The following overview summarizes the reflection coefficients of the four investigated samples. Here, $S_{11}$ (green curves) determines the reflection of the IDTs generating the forward SAWs, while $S_{22}$ (blue curves) corresponds to the backward IDTs. The sharp peaks in $S_{11}$ and $S_{22}$ arise from reflections at the sample edges. Samples 1–3 in (a) – (c) exhibit comparable values for both SAW directions. In contrast, Sample 4 (d) shows overall lower values, however the forward and backward SAWs remain very similar. Thus, the reflection coefficients confirm that the differences in the magnitude of the acousto-electric currents arise from the described contact behavior and are not caused by asymmetries of the SAW delay lines.

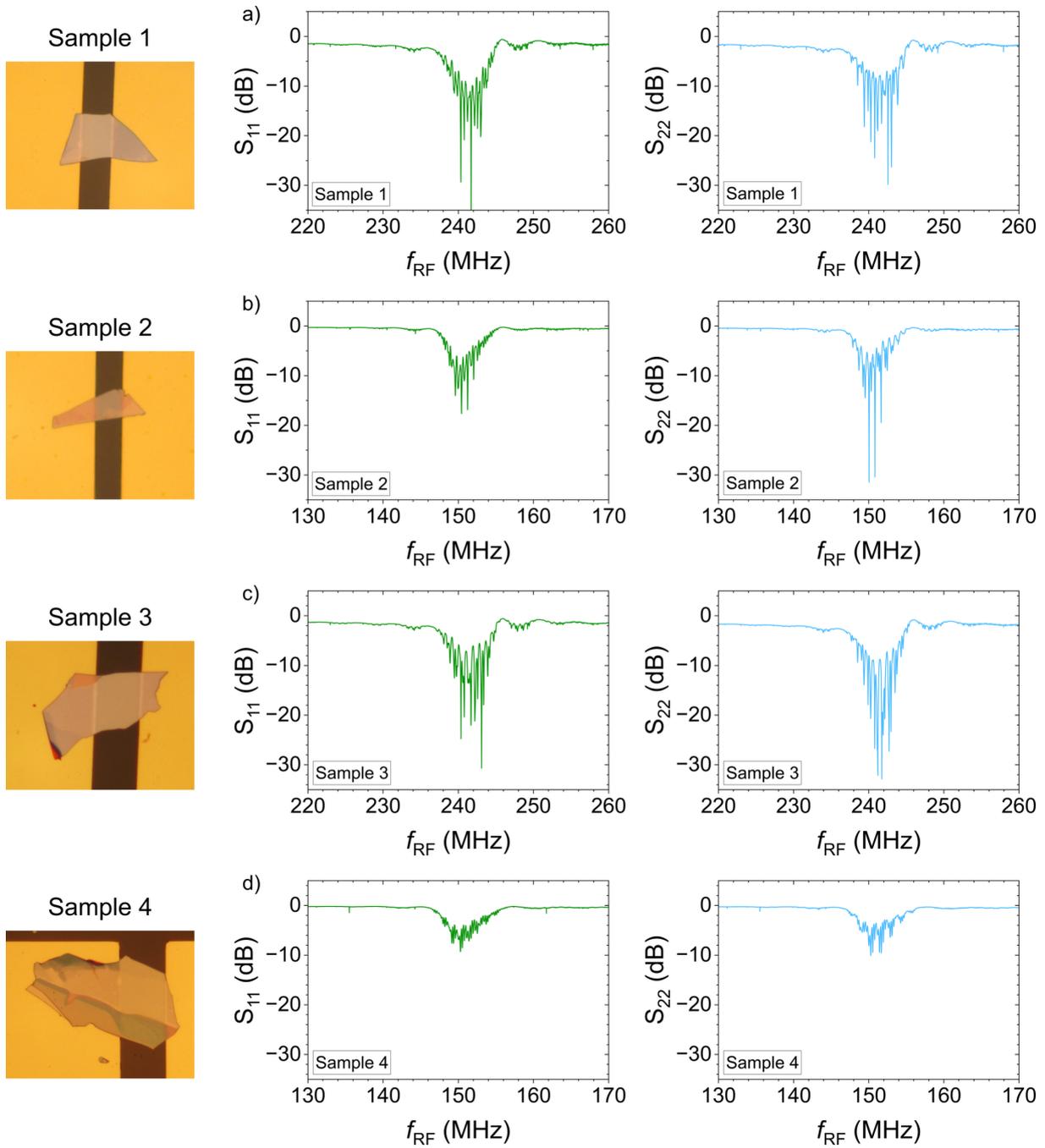

**Figure 9.** Reflection coefficients of Sample 1 (a), Sample 2 (b), Sample 3 (c) and Sample 4 (d).

## 2. Results for Sample 3

The following data show the results for Sample 3. Since the SAW wavelength equals $\lambda_{SAW,3} = 15$ µm, the separation between the electrodes of $d_{SD} = 18$ µm corresponds to more than one acoustic wavelength. Thus, the device functions in the transport regime, as discussed for Sample 1. Similar to Sample 2, Sample 3 also exhibits asymmetric electrical characteristics, which allows the study of the influence of the SAW in the transport regime on asymmetric contacts. Nevertheless, the fundamental results remain consistent with those of Sample 1. All measurements were performed analogously to Samples 1 and 2, as described in the main text.

### 2.1. Pre-characterization

The $WSe_2$ multilayer is placed on top of the two gold electrodes (a). The analysis of the transmission scattering parameter $S_{21}$ confirms a transmission peak at 242MHz with a total bandwidth of $\Delta f_{RF} \approx 8$MHz marked by the purple box (b). The multilayer nature of the $WSe_2$ flake is verified by photoluminescence spectroscopy, revealing a dominant peak at 1.4 eV accompanied by a weak shoulder at 1.6 eV (c).

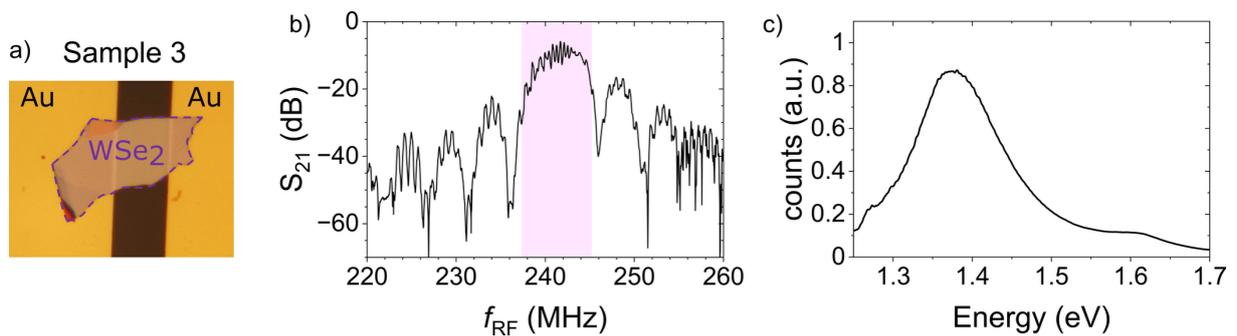

**Figure 10.1.** Pre-characterization – (a) Micrograph of the $WSe_2$ multilayer. (b) Scattering parameter $S_{21}$. (c) Photoluminescence spectroscopy.

## 2.2. Current-voltage ($I_{DS} - V_{DS}$) - characteristics

With values of $I_{DS} = -0.1$ nA to $+10.9$ nA for $V_{DS} = -5$ V and $V_{DS} = +5$ V Sample 3 shows very asymmetric current-voltage characteristics without optical excitation, indicating a very high energy barrier at the drain (a). As expected, this behavior is drastically changed under photodoping of the full flake with a green laser ($\lambda = 532$nm), leading to values of $I_{DS} = -1560$nA to $+1320$ nA for $V_{DS} = -2$ and $V_{DS} = +2$ V (b).

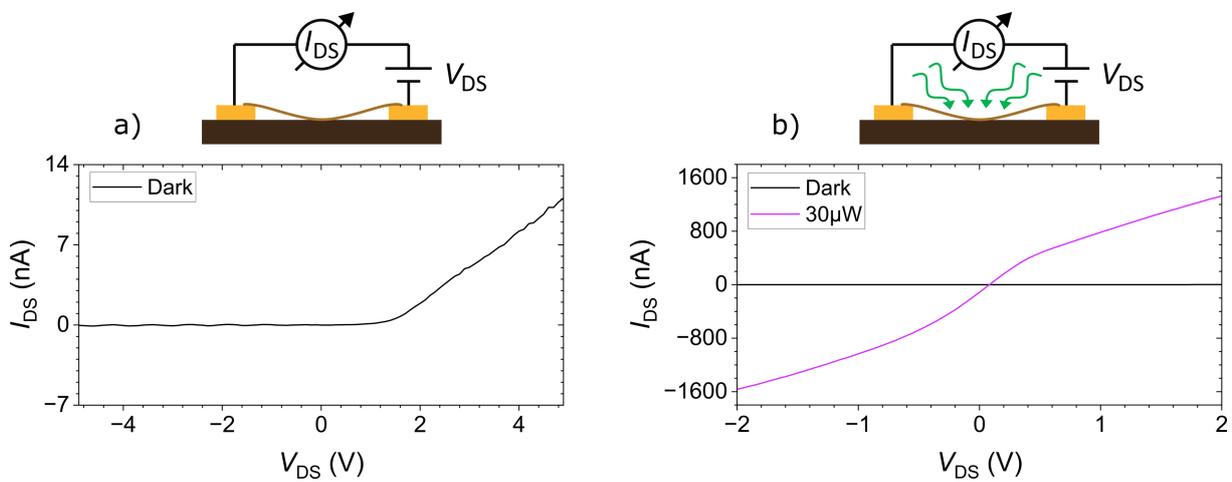

**Figure 2.2.** Current Voltage characterization – (a) ($I_{DS} - V_{DS}$) - characteristics without optical excitation. (b) ($I_{DS} - V_{DS}$) - characteristics under photodoping ($\lambda = 532$nm).

## 2.3. Scanning Photocurrent

The focused laser ($\lambda = 532$nm) is scanned line by line across Sample 3. The induced short-circuit photocurrent (PC) is measured (compare **Figure 3**, Main Text). The false color plot of the PC (a) shows a high negative $I_{DS} = -76$nA at the Drain and weak positive $I_{DS} = +14$nA at the Source electrode, depicted in the horizontal scan at $y = -16$μm (b). These values are in agreement with the asymmetric current-voltage characteristics (**Figure 2.2**) and the acousto-electric current (**Figure 2.4**). The charge carrier dynamics are modeled considering a dominant tunneling barrier and an extended band bending at the drain electrode (c).

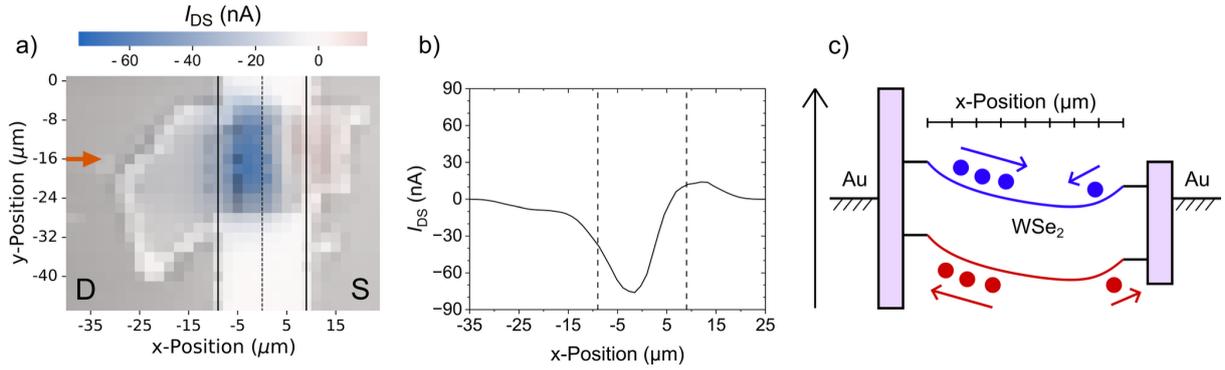

**Figure 2.3.** Scanning Photocurrent – (a) False color plot of the laser position dependent PC with the reflection of the laser in gray scale. (b) Line plot of the PC at $y = -16$μm. (c) Schematic of the energy band/barrier model and induced charge carrier dynamics.

## 2.4. Acousto-electric current

The expected change of the current sign is observed for the two SAW propagation directions in the frequency-dependent acousto-electric currents (AEC) measurement (a). However, the amplitude strongly differs. The forward propagating SAW (green) induces high AEC of up to $I_{DS} = +823$nA while the backward propagating SAW (blue) only leads to a maximum value of $I_{DS} = -8$nA. This is in accordance with the barrier model introduced in Figure 3 since it underpins the effective insertion of holes (weak p-type material) into the drain while the extraction at the source is strongly suppressed. The SAW-Power dependency (b) follows the linear trend shown in the main text for Sample 1 (**Figure 4**a, Main Text).

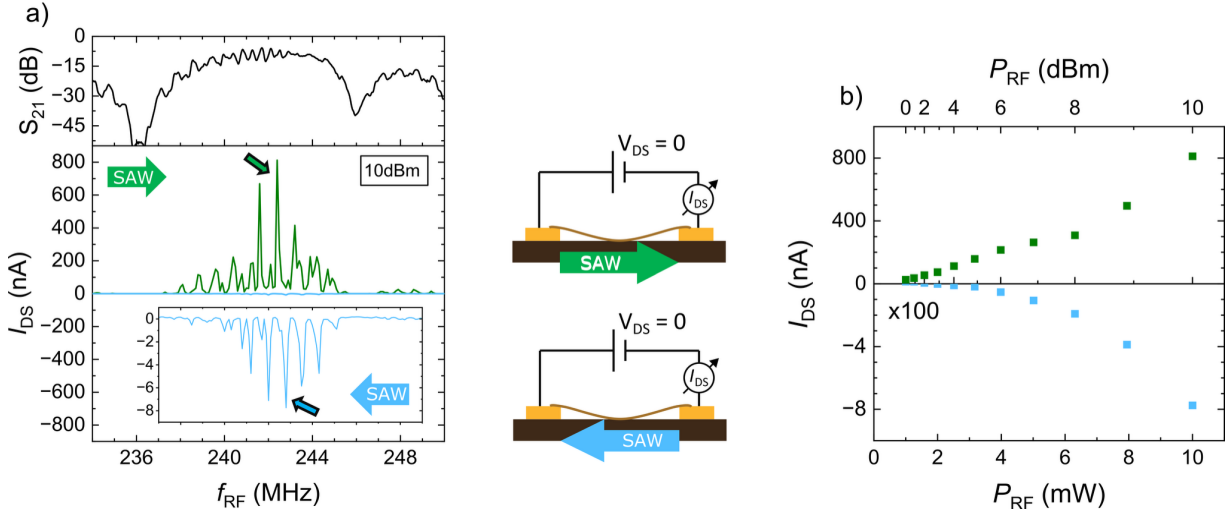

**Figure 2.4.** Acousto-electric current – (a) Upper panel: Scattering parameter $S_{21}$. Lower panel: Acousto-electric current ($I_{DS}$) as a function of the applied SAW frequency. (b) $I_{DS}$ as a function of the SAW power.

## 2.5. Scanning acousto-photoelectric current

The combination of the scanning photocurrent measurement with either a forward or backward SAW (APEC) shows the same fundamental behavior as observed for Sample 1. In the unipolar transport regime at low $P_{RF} \leq -5$ dBm, the negative photocurrent peak (**Figure 2.3**) is shifted from the drain electrode towards the source electrode when applying a backward propagating SAW (a, b). With a value of $I_{DS} = -555$nA at $P_{RF} = -5$dBm we observe an enhancement factor of $\frac{I_{DS}(APEC)}{I_{DS}(PC)} \sim 7$. Vice versa the positive photocurrent peak is located at the drain electrode in case of a forward propagating SAW (e, f) with an enhancement factor of $\frac{I_{DS}(APEC)}{I_{DS}(PC)} \sim 7$ ($I_{DS}$(APEC) = +260nA). At higher $P_{RF}$ = 5dBm and $P_{RF}$ = 10dBm we obtain a vanishing net current due to the bipolar transport of electrons and holes for the backward SAW (c, d). For the forward SAW, however, the transport of intrinsic holes is dominating which leads to a positive current background, independent on the laser position (g, h).

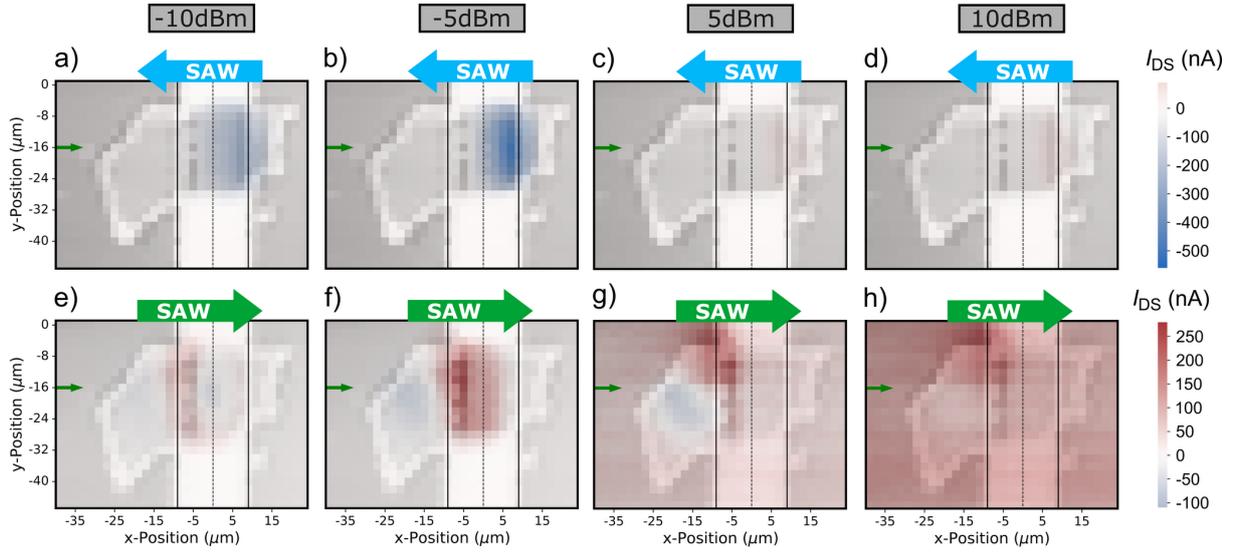

**Figure 2.5.** Scanning acousto-photoelectric current – (a-d) False color plot of the laser position dependent photocurrent for the backward (a-d) and forward (e-h) propagating SAW for applied $P_{RF}$ of -10, -5, 5 and 10dBm.

## 2.6. Charge carrier dynamics

The line plot of the APEC at $y = -16$μm (marked by the corresponding orange arrows in **Figure 2.5**) for low $P_{RF} = -10$dBm and $P_{RF} = -5$dBm exhibits the negative current peaks at the drain electrode for the backward SAW at the drain (a) and the positive peaks at the source electrode for the forward SAW (b). The underlying charge transport mechanisms are consistent with those discussed for Sample 1 in **Figure 7** in the main text.

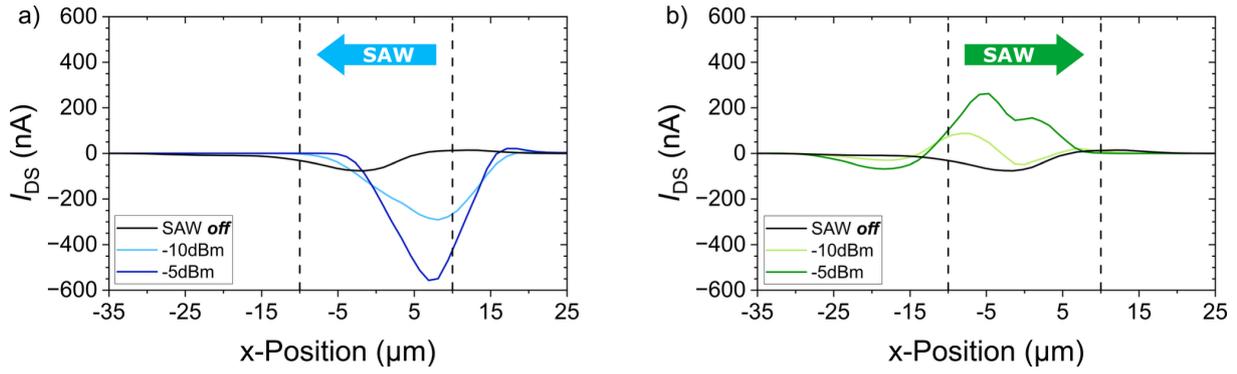

**Figure 2.6.** Line plot of the photocurrent at $y = -16\mu m$ – (a) backward SAW at $P_{RF} = -10$dBm and $P_{RF} = -5$dBm. (b) forward SAW at $P_{RF} = -10$dBm and $P_{RF} = -5$dBm.

## 3. Results for Sample 4

In the last section, we briefly introduce Sample 4. With $\lambda_{SAW.4} = 23$ μm and $d_{DS} = 18$ μm, the ratio between the source–drain gap and the acoustic wavelength (d) is in the same regime as for Sample 1. The electrical contacts exhibit symmetric behavior, as confirmed by the ($I_{DS} - V_{DS}$) - characteristics (a) and the AEC (b). For this reason, all fundamental findings of Sample 4 are consistent with Sample 1. The PC measurement (c) shows a negative peak ($I_{DS} = -31$nA) at the drain electrode and a slightly dominating positive peak ($I_{DS} = +33$nA) at the source electrode, arising from a nearly symmetric distribution of the band bending. The line plots of the APEC at $y = -22$ μm (e), as well as the corresponding false-color plots (f) at $P_{RF} = 5$ dBm (upper panels) and $P_{RF} = 10$ dBm (lower panels), confirm the shift of the positive current peak from source to drain for a forward SAW (left panels) and, conversely, the shift of the negative

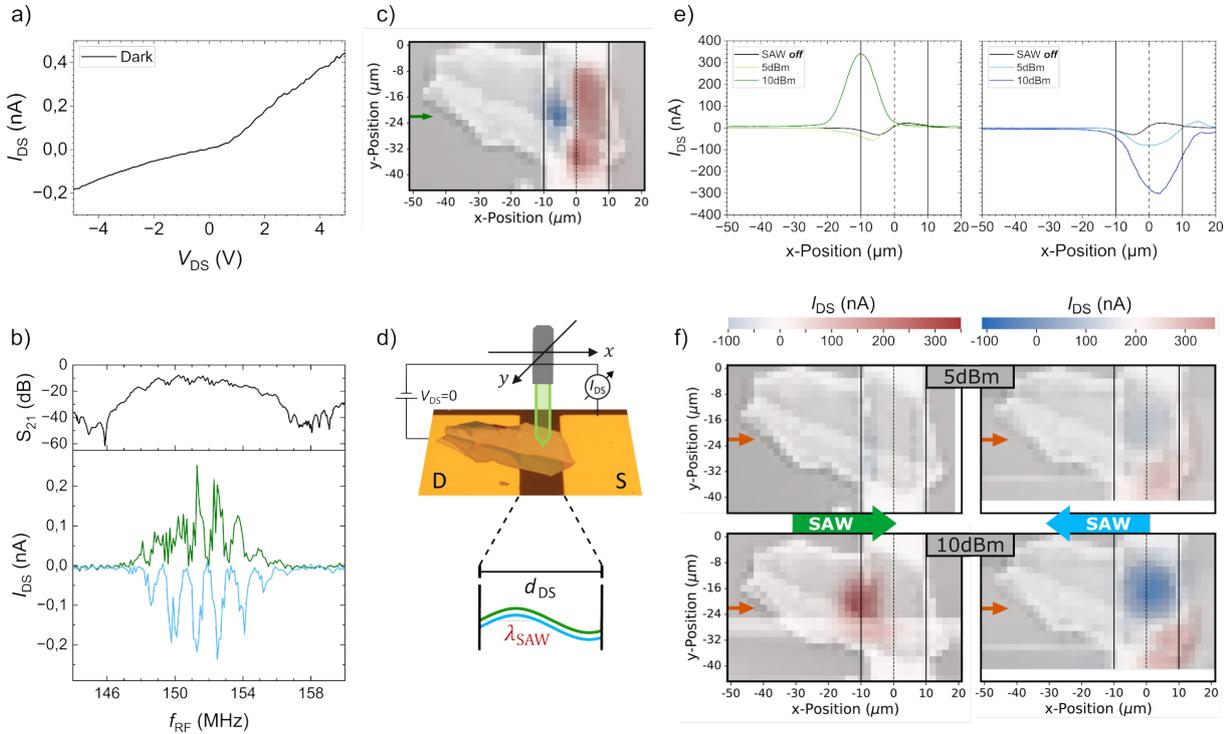

**Figure 3.** Results of Sample 4 – (a) ($I_{DS} - V_{DS}$) – characteristics. (b) AEC. (c) PC. (d) gap to acoustic wavelength ratio. (e) APEC line-plot. (f) APEC false-color plot.

current peak from drain to source for the backward-propagating SAW (right panels).